\documentclass[a4paper, conference]{IEEEtran}

\usepackage{graphicx}
\usepackage{epsfig}
\usepackage[ansinew]{inputenc}
\usepackage{amsmath}
\usepackage{subfig}
\usepackage{epstopdf}
\usepackage{bm}
\usepackage{dblfloatfix}
\usepackage{tabulary}

\usepackage{floatrow}
\begin{document}

%
\title{Application of Machine Learning for Channel based Message Authentication in Mission Critical Machine Type Communication}


\author{\IEEEauthorblockN{Andreas Weinand, Michael Karrenbauer, Raja Sattiraju, Hans D. Schotten}
\IEEEauthorblockA{Institute for Wireless Communication and Navigation\\
University of Kaiserslautern, Germany\\
Email: \{weinand, karrenbauer, sattiraju, schotten\}@eit.uni-kl.de
}
}


%


\maketitle

\begin{abstract}
The design of robust wireless communication systems for industrial applications such as closed loop control processes has been considered manifold recently. Additionally, the ongoing advances in the area of connected mobility have similar or even higher requirements regarding system reliability and availability. Beside unfulfilled reliability requirements, the availability of a system can further be reduced, if it is under attack in the sense of violation of information security goals such as data authenticity or integrity. In order to guarantee the safe operation of an application, a system has at least to be able to detect these attacks. Though there are numerous techniques in the sense of conventional cryptography in order to achieve that goal, these are not always suited for the requirements of the applications mentioned due to resource inefficiency.
In the present work, we show how the goal of message authenticity based on physical layer security (PHYSEC) can be achieved. The main idea for such techniques is to exploit user specific characteristics of the wireless channel, especially in spatial domain. Additionally, we show the performance of our machine learning based approach and compare it with other existing approaches.
\end{abstract}

%
\IEEEpeerreviewmaketitle

\section{Introduction}
\label{intro}
{\let\thefootnote\relax\footnote{This is a preprint, the full paper has been published in Proceedings of 23th European Wireless Conference (European Wireless 2017), \copyright 2017 IEEE. Personal use of this material is permitted. However, permission to use this material for any other purposes must be obtained from the IEEE by sending a request to pubs-permissions@ieee.org.}}
\IEEEpubidadjcol

Currently, there are two major developments in the area of wireless systems regarding highly safety critical applications such as closed loop control applications in the area of industrial systems (\cite{Bockelmann.2017}) and network based autonomous driving functions in the area of vehicular systems. These have much higher requirements regarding reliability, availability and also latency compared to common applications, such as e.g. media streaming or web browsing over IEEE 802.11 based wireless systems or today's cellular 4G LTE networks. Another important requirement in the area of mission critical machine type communication (MC-MTC) is the fact, that secure transmission of data has to be taken into account. Due to the sensitive information transmitted in the mentioned scenarios, it is necessary to guarantee a high degree of information security. Especially confidentiality, authenticity, as well as integrity of the transmitted data has to be ensured to prohibit a wide range of possible passive and active cyber attacks. For this purpose, encryption of transmitted messages is necessary in order to be sure of the confidentiality of the respective payload and authentication of the originator as well as integrity checking of received messages have to be ensured before they are consumed and processed by a safety critical application.
%

Although there are conventional cryptography techniques to ensure authenticity as well as integrity of message payload, these require a not negligible amount of resources. Especially they lead to increase in message size due to the fact that, for example message authentication codes (MAC), which are commonly used in deployed systems (e.g. in IEEE 802.15.4 based systems), add a cryptographic check sum to the actual message payload. The recommendation of the Internet Engineering Task Force is to either use a CMAC (cipher based MAC), e.g. based on a AES-128 block cipher, or a HMAC which is based on a cryptographic hash function. For AES-128 based CMAC a maximum shortening to 64 Bit is recommended \cite{RFC_CMAC.2006}, while for HMAC a minimum MAC size of 80 Bit is recommended \cite{RFC_HMAC.1997}. If we now assume that the payload of a MC-MTC packet in the uplink has a length of 400 Bit (the dimension of this assumption is e.g. confirmed by \cite{Bockelmann.2017} and \cite{Osman.2015}), then the overhead and with this the additional latency introduced in case of a AES-128 based CMAC of $64$ Bit length would be $13.8\%$. Another important issue is, that key based schemes such as MAC are only able to protect the message payload from the mentioned attacks. An attacker is still able to perform attacks such as address spoofing, or even worse, record a message and replay it after a while (if counters are used to prohibit that, additional latency is added as well). Due to these drawbacks, another idea is to check for message authenticity at a lower level by taking physical properties of the radio link signal into account. In this work, we focus on a keyless approach in order to achieve that goal based on estimating the wireless channel and detection of different users based on that. As mentioned, MC-MTC is considered here, for which it is assumed that frequent and periodic data transmissions (e.g. once per ms \cite{Bockelmann.2017}, \cite{Osman.2015}) and with this channel estimation at the same rate is available. For experimental evaluation, we consider an OFDM system and based on the respective frequency domain channel estimations, we decide from which source a received data packet was transmitted.\\
The remainder of the work is organized as follows. In section \ref{related work} we give a short overview on related work with respect to previous considered approaches and in section \ref{system_model_section} we describe the system model. The machine learning approach for physical layer based message authentication is presented in section \ref{approach}. In section \ref{results} we present the results of our work in form of an experimental evaluation of the considered approach and section \ref{CONC} finally concludes the paper. 

\section{Related Work}
\label{related work}
Several approaches on exploiting the wireless channel for security purposes, also known as PHYSEC, have been investigated recently. In \cite{Jorswieck.2015} a good overview on this topic is given. While many works have focused on extracting secret keys between two communicating devices, such as \cite{Guillaume.}, \cite{Zenger.2014}, \cite{Ambekar.2012b}, the focus of our work is on guaranteeing secure transmission with respect to authenticity of data packets from one device to another. One of the first works considering that idea has been for example \cite{Xiao.2007}, where an approach based on channel measurements and hypothesis testing is presented for static scenarios and is later in \cite{Xiao.2008} extended to time-variant scenarios. Though the results of these works are plausible, they are obtained from simulations of the wireless channel only. Therefore, in our work we will focus on real world channels. In \cite{Pei.2014}, two approaches based on machine learning, Support Vector Machine and Linear Fisher Discriminant Analysis, are presented. The approach considered in \cite{Tugnait.2010} is similar to our approach, as they propose a CSI-based authentication method. However, they consider a single carrier system, while we will focus on a multi carrier system in our work. The second approach considered in \cite{Tugnait.2010} is whiteness of residuals testing. In \cite{Shi.2013} an RSSI-based approach for body area networks is presented. The work in \cite{Refaey.2014} considers a multilayer approach based on OFDM to guarantee authentication of TCP packets. A Gaussian Mixture Model based technique in combination with exploitation of the channel responses for different antenna modes is considered in \cite{Gulati.2013}. In \cite{Baracca.2012} a MIMO OFDM system is considered and a generalized likelihood ratio test is used as detection method. Further, the authors derive information theoretic bounds for the channel based detection problem. The authors in \cite{Caparra.2016} consider a cellular Internet-of-Things system and propose a scheme for channel based message authentication of nodes with the help of anchor nodes, which are assumed to have set up trust to the concentrator node before.

\section{System Model}
\label{system_model_section}
In this section we describe the system model and the channel model considered in our work. Further, the attacker model is introduced.
\subsection{Communication Model}
\begin{figure}[h]
\centering
\includegraphics[width=0.9\textwidth]{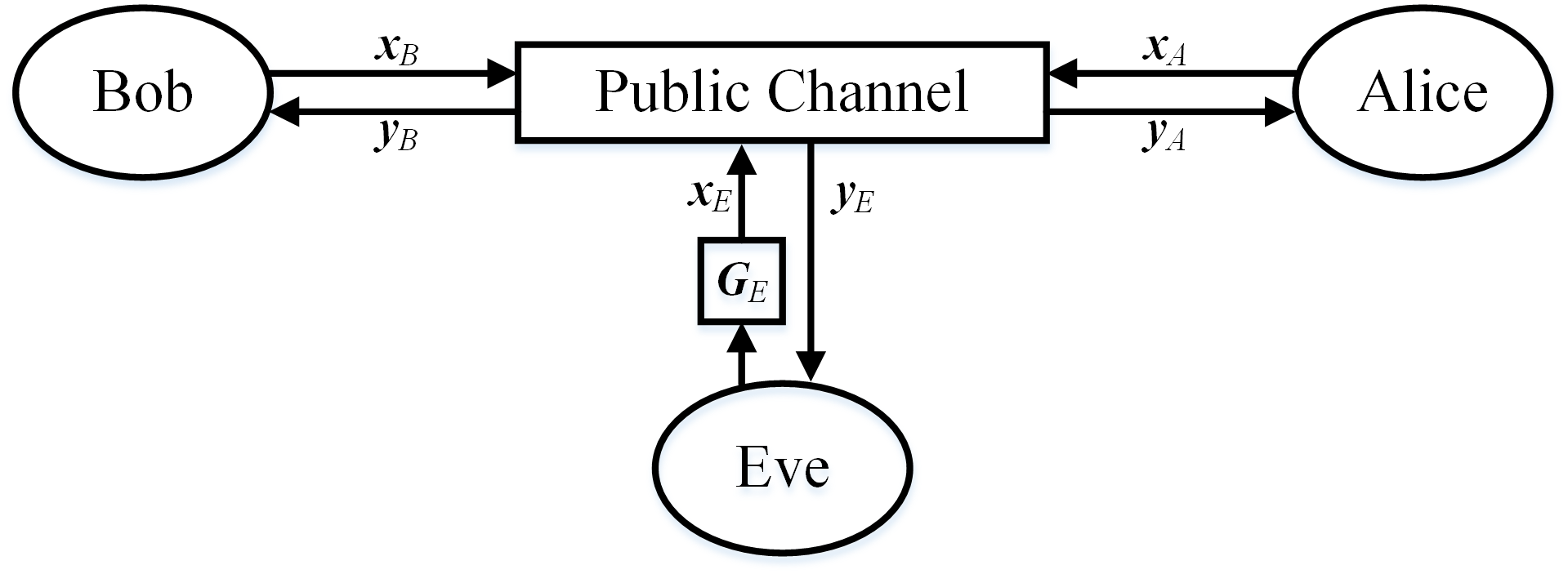}
\caption{System Model}
\label{system_model}
\end{figure}

We consider two users, Alice and Bob, who want to exchange authenticated messages with each other over a public channel. A third user Eve, who is an adversarial user, tries to inject illegal messages masqueraded as one of the legal users Alice or Bob. Fig. \ref{system_model} shows the possible message flows for all considered users. Messages transmitted by users are denoted with $\boldsymbol{x}_u$ and messages received by users with $\boldsymbol{y}_u$, where $u$ denotes user Alice ($u=A$), Bob ($u=B$) or Eve ($u=E$).

\subsection{Channel Model and Channel Estimation}
In our work, we focus on channel state information (CSI), which is e.g. computed at the receiver side in form of a frequency domain channel estimation. The main idea for all PHYSEC based techniques is to make use of the fast spatial decorrelation property of the wireless channel. For that purpose, the users need to be able estimate the channel to each other. The received signal at a user $u_i$ is
\begin{equation}
\boldsymbol{y}_{u_i}=\boldsymbol{H}_{u_iu_j}{\boldsymbol{x}_{u_j}}+{\boldsymbol{n}_{u_iu_j}}
\label{eq:}
\end{equation}
with
\begin{equation}
\boldsymbol{H}_{u_iu_j}=[{h}^0_{u_iu_j},\ldots, {h}^{M-1}_{u_iu_j}]
\label{eq:}
\end{equation}
being the $M$-dimensional channel matrix of the frequency-selective SISO fading channel between user $u_i$ and $u_j$ (transmitter) and ${h}^\ell_{u_iu_j}$ being the gain of the $\ell$-th sample in frequency domain ($\ell=0,\ldots,M-1$). It can be estimated as 
\begin{equation}
\boldsymbol{\hat{H}}_{u_iu_j}=\boldsymbol{H}_{u_iu_j}+\boldsymbol{\epsilon}_{u_iu_j}
\label{eq:}
\end{equation}
by user $u_i$. Due to noise $\boldsymbol{n}_{u_iu_j}$ which is modelled as a gaussian random variable with zero mean and variance $\boldsymbol{\sigma^2}_{\boldsymbol{n}_{u_iu_j}}$ the channel estimation is not perfect and errors $\boldsymbol{\epsilon}_{u_iu_j}$ occur. 

\subsection{Attacker Model}
A typical scenario for an attacker Eve is, that he is at a spatially different location compared to Bob and Alice (we assume a distance of more than the wavelength of the transmitted signal respectively) and uses advanced equipment such as directed antennas and high sensitivity receivers in order to maximize the coverage to his benefit. We also assume perfect knowledge of the underlying communication protocol at Eve to run active attacks such as masquerade attacks, replay attacks or address spoofing attacks by introducing messages $\boldsymbol{x}_{\mbox{\scriptsize E}}$. It is further assumed, that Eve is able to eavesdrop on ongoing transmissions of Alice ($\boldsymbol{y}_{\mbox{\scriptsize E}}=\boldsymbol{H}_{\mbox{\scriptsize EA}}{\boldsymbol{x}_{\mbox{\scriptsize A}}}+\boldsymbol{n}_{\mbox{\scriptsize EA}}$) and Bob ($\boldsymbol{y}_{\mbox{\scriptsize E}}=\boldsymbol{H}_{\mbox{\scriptsize EB}}{\boldsymbol{x}_{\mbox{\scriptsize B}}}+\boldsymbol{n}_{\mbox{\scriptsize EB}}$) and estimate the respective channels $\boldsymbol{H}_{\mbox{\scriptsize EA}}$ and $\boldsymbol{H}_{\mbox{\scriptsize EB}}$. With this information, Eve is able to manipulate its injected messages $\boldsymbol{x}_{\mbox{\scriptsize E}}$ with a pre-processing filter $\boldsymbol{G}_{\mbox{\scriptsize E}}$ in such a way, that the channel between the legal users is imitated. For this work we assume that $\boldsymbol{G}_{\mbox{\scriptsize E}}=1$ and with this $\boldsymbol{H}_{\mbox{\scriptsize AE}} \ne \boldsymbol{H}_{\mbox{\scriptsize AB}}$ and $\boldsymbol{H}_{\mbox{\scriptsize BE}} \ne \boldsymbol{H}_{\mbox{\scriptsize BA}}$. It is not assumed that Eve is gaining physical access to Alice or Bob to accomplish invasive attacks such as hardware modification. Further, other active attacks such as Denial-of-Service attacks due to jamming are not considered as well. It is assumed, that the legal communicating participants Bob and Alice have already carried out initial user authentication to each other and have set up trust in a secure way. Attacks on the initial authentication stage are not considered.

\section{Machine Learning and Channel based Authentication}
\label{approach}
\begin{figure}[h]
\centering
\includegraphics[width=0.35\textwidth]{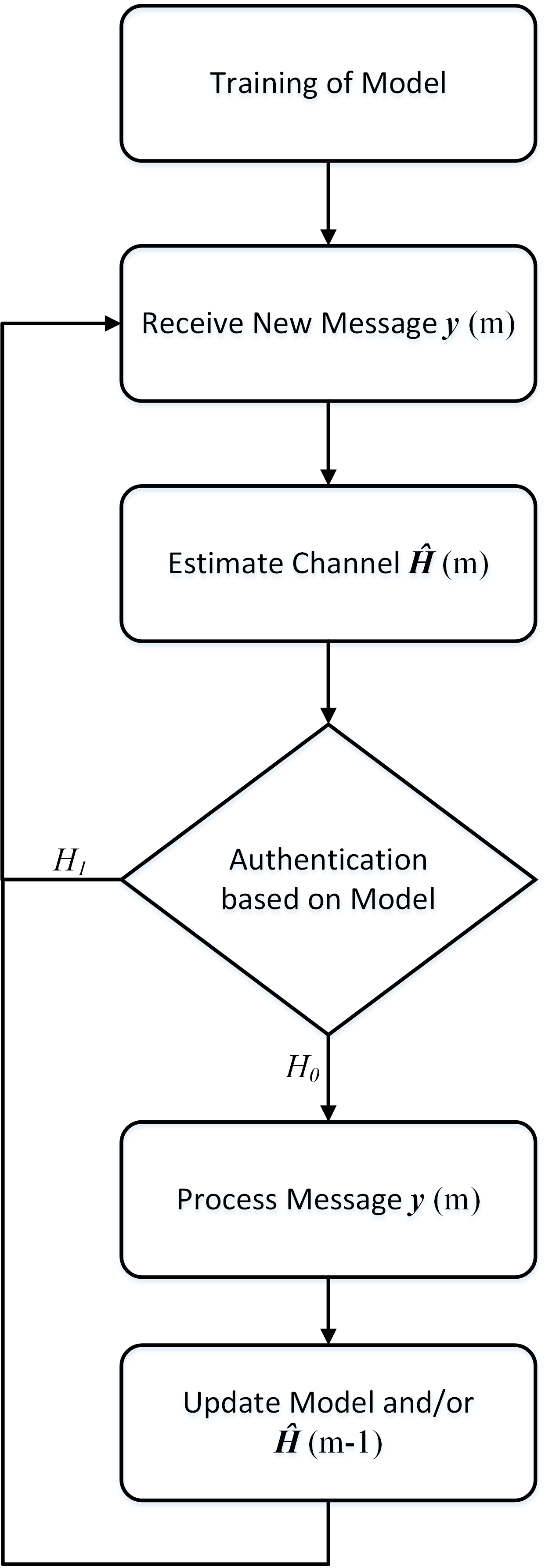}
\caption{Channel based Message Authentication using Machine Learning}
\label{ML_flow_diag}
\end{figure}

In this section, we show how channel based message authentication with the help of machine learning can be achieved. In general, all machine learning algorithms contain two stages. In the first stage, the respective algorithm needs to be trained to the kind of data it is intended to estimate. In our case, this data are channel estimations taken on received data packets. If we now consider the users Alice and Bob again, e.g. Bob wants to transmit authenticated messages to Alice. Alice will estimate the channel as $\boldsymbol{\hat{H}}_{\mbox{\scriptsize AB}}{(k)}$ during the training phase for each of the $T$ received messages $\boldsymbol{y}_{\mbox{\scriptsize A}}{(k)}, k=0,\ldots,T-1$. In this stage we also assume, that the acquired data is labeled. This means that the messages $\boldsymbol{x}_{\mbox{\scriptsize B}}$ sent by Bob contain cryptographic information from higher layers (e.g. certificates) in order to authenticate to Alice initially during that stage. The acquired data $\boldsymbol{\hat{H}}_{\mbox{\scriptsize AB}}{(k)}$ and the respective labels are fed to the training procedure of the machine learning algorithm. In the second stage, the trained model is used in order to make a decision on further received messages $\boldsymbol{y}_{\mbox{\scriptsize A}}{(m)}$ at time $m$. If Alice estimates the channel now as $\boldsymbol{\hat{H}}{(m)}$, there are two hypothesis, either
\begin{equation} \label{eq1}
\begin{split}
\mathcal{H}_0&: \boldsymbol{y}_{\mbox{\scriptsize A}}{(m)} \ \rm was\ sent\ by\ Bob,\ or\ \\
\mathcal{H}_1&: \boldsymbol{y}_{\mbox{\scriptsize A}}{(m)} \ \rm was\ not\ sent\ by\ Bob.\
\end{split}
\end{equation}
In case of hypothesis $\mathcal{H}_0$ 
\begin{equation}
\boldsymbol{\hat{H}}{(m)}=\boldsymbol{H}_{\mbox{\scriptsize AB}}{(m)}+\boldsymbol{\epsilon}_{\mbox{\scriptsize AB}}{(m)},
\label{eq:}
\end{equation}  
and in case of hypothesis $\mathcal{H}_1$ 
\begin{equation}
\boldsymbol{\hat{H}}{(m)}\ne\boldsymbol{H}_{\mbox{\scriptsize AB}}{(m)}+\boldsymbol{\epsilon}_{\mbox{\scriptsize AB}}{(m)}.
\label{eq:}
\end{equation}
Due to temporal variations in the channel $\boldsymbol{H}_{\mbox{\scriptsize AB}}$, these need to be somehow considered in the second stage. Otherwise it is not possible to distinguish between channel variations and messages introduced by an attacker anymore at some point. There are two possibilities in order to achieve that goal, either the machine learning algorithm needs to be updated regularly, or instead of this, the difference of the current channel estimate to the previous one 
\begin{equation}
\Delta\boldsymbol{\hat{H}}{(m)}=\boldsymbol{\hat{H}}{(m)}-\boldsymbol{\hat{H}}{(m-1)}
\label{eq:}
\end{equation}
is used as input data for the machine learning algorithm. In the first case, the model update interval should be below the channel coherence time $T_c$ of the $\boldsymbol{H}_{\mbox{\scriptsize AB}}$ channel in order to catch up with these variations. Whereas in both cases, the temporal difference between two subsequent channel estimations should always be below $T_c$. Fig. \ref{ML_flow_diag} shows all steps of the general procedure described above.

Every machine learning estimator has basically two main performance indicators, the detection probability $P_{\mbox{\scriptsize D}}$ and the false alarm rate $P_{\mbox{\scriptsize FA}}$. In our case, the detection probability 
\begin{equation}
P_{\mbox{\scriptsize D}}=p(\mathcal{H}_1|\boldsymbol{\hat{H}}{(m)} \mathrm{\ due\ to\ } \boldsymbol{x}_{\mbox{\scriptsize E}})
\label{eq:}
\end{equation}
denotes the probability of detecting Eve as the transmitter of $\boldsymbol{y}_{\mbox{\scriptsize A}}{(m)}$ under the condition that it was truly sent by Eve and the false alarm rate
\begin{equation}
P_{\mbox{\scriptsize FA}}=p(\mathcal{H}_1|\boldsymbol{\hat{H}}{(m)} \mathrm{\ due\ to\ } \boldsymbol{x}_{\mbox{\scriptsize B}})
\label{eq:}
\end{equation}
denotes the probability of detecting Eve as the transmitter of $\boldsymbol{y}_{\mbox{\scriptsize A}}{(m)}$ under the condition that it was truly sent by Bob. Both performance metrics need to be considered jointly, as they can not be optimized independently. The relation of both can be shown in a receiver operating characteristic (ROC) curve.

\section{Results}
\label{results}






In this section we describe our setup for the experimental evaluation and show the results of our work. Further, we compare them with results of related work.
\subsection{Experimental setup}
To evaluate our concepts, we use USRP N210 SDR platforms from Ettus Research with SBX daughterboards. We use GNURadio OFDM transmitter and receiver blocks to process data packets and perform channel estimation on each received data packet. A setup with an FFT size of $64$ is considered and $48$ active subcarriers. The cyclic prefix length is $16$ samples at a baseband sample rate of $3.125$ MSps, whereas the carrier frequency is $2.45$ GHz. For each received data packet, the initial channel taps are calculated based on the known Schmidl and Cox preamble \cite{Schmidl.1997} which is also used to calculate the frequency offset at the receiver (actually this preamble consists of two OFDM symbols). In each message, this preamble is followed by $37$ data symbols yielding a time resolution of 998.4 $\mu$s for the channel estimations. As a first step, we consider a static setup where all participants do not move during transmitting and receiving. The environment is a mixed office/lab area with a lot of objects and metal walls. Due to this we assume that at least some amount of multipath propagation is existing and with this frequency selective channels. We record data for several different locations of Bob and Eve respectively, yielding multiple different constellations of \mbox{Bob-Alice} and \mbox{Eve-Alice} pairs. 

We choose Gaussian Mixture Model (GMM) based clustering as detection method in order to authenticate received messages. We considered a block size of $N=1000$ data sets in order to update the GMM, whereas one block is used for training the model and $99$ blocks are used in order to test it. The attack intensity was kept at $50\%$. We use the magnitude of normalized channel estimations 
\begin{equation}
\boldsymbol{{F}}(m)=\frac{\big|\boldsymbol{\hat{H}}(m)\big|}{\sum\limits_{\ell=0}^{M-1}\Big|{\hat{h}}^{\ell}(m)\Big|}
\label{eq:}
\end{equation}
as input data (feature) for the GMM estimator.

\begin{figure}[t!]
\centering
\includegraphics[width=\textwidth]{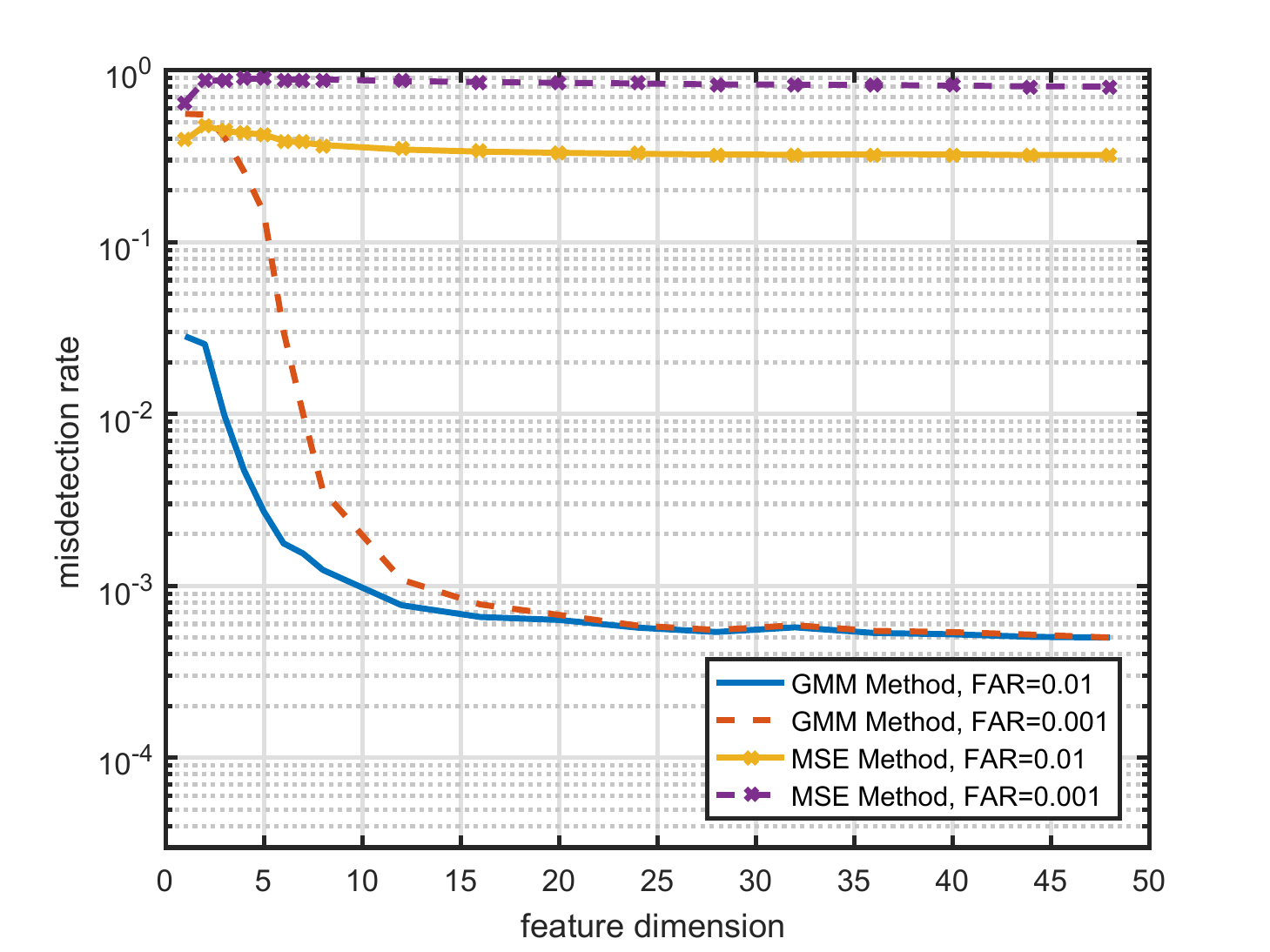}
\caption{Misdetection rate for different false alarm rates}
\label{misdetection}
\end{figure}

\subsection{Performance of GMM based Clustering}
Fig. \ref{misdetection} shows the probability of misdetection  
\begin{equation}
P_{\mbox{\scriptsize MD}}=1-P_{\mbox{\scriptsize D}}
\label{eq:}
\end{equation}
or misdetection rate (MDR) in dependence of the number of estimated subcarriers $M$ used for the detection.
The influence of the accuracy of the channel estimation in the sense of different dimensions of the feature vector $\boldsymbol{{F}}(m)$ can be seen, as for increasing values of $M$ the MDR overall drops. If only $M=4$ subcarriers (equally spaced) are considered, the MDR of the GMM method is at $25.68\%$ at a false alarm rate of $0.1\%$ and at $0.47\%$ at a false alarm rate of $1\%$. For the case of $M=16$ subcarriers (equally spaced) considered, the MDR decreases by $99.67\%$ to $0.08\%$ at a false alarm rate of $0.1\%$ and by $85.11\%$ to $0.07\%$ at a false alarm rate of $1\%$. In case of the MSE based detection method from \cite{Weinand.2016}, the MDR is at $43.03\%$ in case of $M=4$ and $1\%$ false alarm rate, which is an increase of $9055.32\%$ compared to the GMM method and in case of a false alarm rate of $0.1\%$, the MDR is at $89.21\%$, which is an increase of $247.39\%$ compared to the GMM method. In case of $M=16$, the MDR of the MSE based method is at $33.62\%$ at a false alarm rate of $1\%$, which is an increase of $479.29\%$ compared to the GMM method. For a false alarm of $0.1\%$ the MDR is at $85.36\%$, which is an increase of $1066\%$ compared to the GMM method.

\begin{figure}[b!]
\centering
\includegraphics[width=\textwidth]{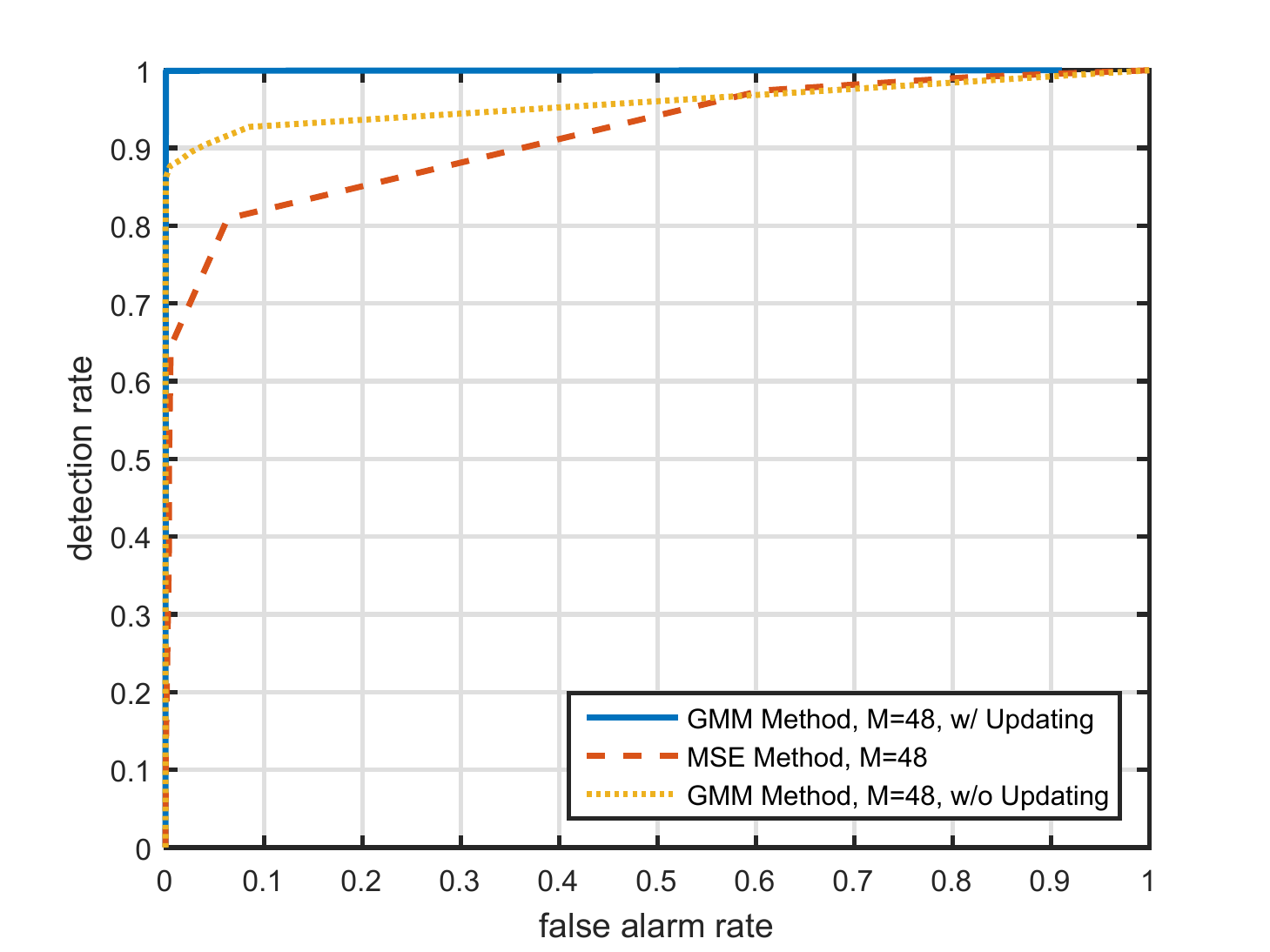}
\caption{GMM based method with und without updating}
\label{GMM_update}
\end{figure}

Fig. \ref{GMM_update} shows the influence of the updating process within the GMM based clustering method. If the updating process of the model is switched off, the detection rate decreases significantly. At a false alarm rate of $1\%$, the detection rate for GMM based detection without updating is at $88.06\%$, which is a performance loss of $11.9\%$ compared to the GMM based detection with updating. However, the GMM based method without updating still outperforms the MSE based method ($65.78\%$ detection rate) by a $33.87\%$ higher detection rate. For a false alarm rate of $0.1\%$, the MSE method performance loss is even worse as the detection rate decreases to $17.69\%$ which is $79.53\%$ less compared to the GMM method without updating ($86.43\%$ detection rate) and $82.3\%$ less compared to the GMM method with updating ($99.96\%$ detection rate). The performance loss for the GMM method without updating compared to the GMM method with updating is $13.54\%$.

\subsection{Discussion of Results}
The experimental evaluation of our method shows us basically two things. The first one is, that the performance of our GMM based clustering method increases as the dimension of the considered feature, which is in our case the number of equally spaced estimated frequency points $M$ for each received message, also increases (with this the misdetection rate decreases). However, if the model that classifies the input data is not updated (or a method that does not consider updating of the classifier during operation is used), the performance is decreased, which leads to the conclusion that only classifiers which allow for online updating fulfill the needs for the channel based authentication approach.\\ 
\begin{table}[t!]
\caption{Performance of related work (selection)}
\label{table_example}
\footnotesize
\centering
\begin{tabulary}{1.0\textwidth}{|c|c|c|c|c|}
\hline
Work & Detection Method & Feature & max. $P_{\mbox{\tiny D}}$ & min. $P_{\mbox{\tiny FA}}$\\
\hline
\cite{Gulati.2013} & GMM & Antenna Mode & $99.9\%$ & $0.4\%$ \\
\hline
\cite{Pei.2014} & SVM, LFDA & RSSI, TOA, CC & $95\%$ & $1\%$ \\
\hline
\cite{Tugnait.2010} & NPHT & CIR (TD) & $99.9\%$ & $1\%$ \\
\hline
\cite{Xiao.2008} & NPHT & CIR (FD) & $99\%$ & $1\%$ \\
\hline
Here & GMM & CIR (FD) & $99.96\%$ & $0.1\%$ \\
\hline
\end{tabulary}
\end{table}
If we compare our GMM based approach with related ones (see Tab. \ref{table_example}), the performance of it is similar to them, though some of them use different features that are fed to the detection method (which as well differs from our detection method in some cases). Overall, our approach performs comparable to the approaches of the related work, and compared to e.g. \cite{Pei.2014} even slightly better for the detection rate ($5.22\%$ higher) and in case of the false alarm rate ($90\%$ lower), the performance is better as well. 

\section{Conclusion and Future Work}
\label{CONC}
Our proposed method of taking user specific channel characteristics in form of the frequency domain OFDM channel estimation into account in order to identify and authenticate them seems to be a promising technique in order to achieve that goal in a very efficient way. Especially for MC-MTC messages, the application of channel based message authentication can lead to reduced transmission latency. Further, the combination of both, MC-MTC and channel based authentication is essential considering system efficiency, as both rely on frequent transmission of data packets. With this, frequent channel estimations are available, which can be reused with little to no further effort.
Though the maximum achieved detection rate is high at $99.96\%$, the method needs still to be improved in order to get more reliable decisions. For that purpose, additional features within the detection process such as RSSI or TOA might be helpful. Further, the tradeoff between complexity and performance of the detection method is an issue that needs to be addressed in future work. Therefore additional detection methods could be investigated. To gain robustness due to errors in channel estimation induced by noise, approaches such as in \cite{Ambekar.2014} might be suited in order to reduce this effect based on pre-processing of channel estimates. Additionally we also want to focus on a mobile setup with little to moderate velocities in order to verify that the method also works well under these conditions.

\section*{Acknowledgment}
A part of this work has been supported by the Federal Ministry of Education and Research of the Federal Republic of Germany (BMBF) in the framework of the project 16KIS0267 HiFlecs. The authors would like to acknowledge the contributions of their colleagues, although the authors alone are responsible for the content of the paper which does not necessarily represent the project. 


\bibliographystyle{IEEEtran}
\bibliography{references_EW2017}
%

%
%
%

\end{document}